\documentclass[12pt]{JHEP3}
\usepackage{amssymb}
\usepackage{amsfonts}
\usepackage{epsf}

\font\mybb=msbm10 at 11pt
\def\bb#1{\hbox{\mybb#1}}

\def\bR {\bb{R}}

\renewcommand{\a}{\alpha}
\renewcommand{\b}{\beta}

\newcommand{\rmd}{{\rm d}}
\newcommand{\m}{\mu}
\newcommand{\n}{\nu}

\def\be{\begin{equation}}
\def\ee{\end{equation}}
\def\bea{\begin{eqnarray}}
\def\eea{\end{eqnarray}}
\def\ba{\begin{array}}
\def\ea{\end{array}}
\def\bi{\begin{itemize}}
\def\ei{\end{itemize}}

\title{{\bf Holography of the {\cal N}=1 Higher-Spin Theory on AdS$_4$}}

\author{Robert G. Leigh\\
        Department of Physics\\
        University of Illinois at Urbana-Champaign\\
        Urbana, IL 61801, USA\\
{\rm and}\\ CERN Theory Division, \\
CH-1211 Geneva 23,
Switzerland\\
  Email: \email{rgleigh@uiuc.edu}}
\author{Anastasios C. Petkou\\
CERN Theory Division, \\
CH-1211 Geneva 23,
Switzerland\\
Email: \email{tassos.petkou@cern.ch}
}

\abstract{We argue that the ${\cal N}=1$ higher-spin theory on AdS$_4$ is
holographically dual
to the ${\cal N}=1$ supersymmetric critical $O(N)$ vector model in
three dimensions. This appears to be a special form of the AdS/CFT correspondence in
which both regular and irregular bulk modes have similar roles and their
interplay leads simultaneously to both the free and the 
interacting phases of the boundary theory. We study various boundary
conditions that correspond to boundary deformations connecting, for
large-$N$, the free and interacting boundary theories. We point out the
importance of parity in this holography and elucidate the Higgs
mechanism responsible for the breaking of higher-spin symmetry for
subleading $N$.
}
\preprint{CERN-TH/2003-095\\
hep-th/0304217}

\keywords{AdS-CFT Correspondence, Conformal Field Theory, Supersymmetry, Duality}
\begin{document}

\section{Introduction and the Klebanov-Polyakov proposal}

It has been long understood that consistent higher-spin gauge theories
admit AdS spacetimes as vacua \cite{Vasiliev}. Nevertheless, only recently the question of
the holography of higher-spin theories has been raised \cite{HMS}. The
interest in this holography currently is growing as it is
gradually realized that it touches upon important issues such as the
small tension limit of string theory and the holography of weakly
coupled quantum field theories. (some other recent work includes
\cite{HSrecent}.)

A concrete proposal for the holographic dual of a simple higher-spin
theory was made in \cite{KP}. Consider the minimal bosonic higher-spin
algebra in $d=4$ 
\be
\label{minbHS}
hs(4)\ni SO(3,2)\,.
\ee
The unitary irreducible representations (UIR) of $Spin(3,2)$ are
characterized by the quantum numbers of the subgroup
$SO(2,1)\times SO(1,1)$; they are labeled $D(\Delta,s)$, with $\Delta$ the
dimension and $s$ the total spin. The massless UIR's saturate the
unitarity bound $\Delta\geq s+1$. (In
addition, there are the exceptional UIR's $D(2,0)$, $D(1/2,0)$ and
$D(1,1/2)$. The latter two are singletons.) The ``currents'' that can be
obtained as symmetric composites of the basic scalar singleton UIR
$D(1/2,0)$ have even spin $s=0,2,4,...$. This is the spectrum of the
minimal bosonic higher-spin theory
\begin{equation}\label{eq:onezerotower}
\left[D(1/2,0)\otimes D(1/2,0)\right]_S=D(1,0)\oplus\sum_{s=1}^{\infty} D(2s+1,2s)\,.
\end{equation}

On $AdS_4$, the realization of this minimal higher-spin bosonic theory
may be consistently constructed, although only partial information is
explicitly available for the action of the theory. For each state in the
spectrum, one considers a corresponding $AdS_4$ field. In particular,
the $D(1,0)$ UIR is associated to a conformally coupled scalar on $AdS_4$
\be
\label{confscalar}
I_4 =\frac{1}{2\kappa_4^2}\int\rmd^4x\sqrt{-g}\left(-\frac{1}{2}(\partial h)^2
-\frac{1}{2}m^2 h^2 +\cdots\right)\,,
\ee
with $m^2 =-2$. Similar, yet largely unknown, terms should be written
for the higher-spin UIR's.

The spectrum of the higher-spins is by construction in one-to-one
correspondence with the spectrum of the conserved higher-spin currents
in a free bosonic theory in $d=3$. To study the holographic dual of
(\ref{confscalar}) one can always normalize the 2-pt functions in the
boundary to be of order 1, and then observe that the $n$-pt correlation
functions for $n\geq 3$ are proportional to $(2\kappa_4^2)^{n/2}$, i.e.,
are suppressed by powers of the Planck length in AdS$_4$. This may be
taken to imply that the elementary fields in the holographic dual of the
minimal bosonic higher-spin theory carry a certain group representation.
 It was observed in \cite{KP} that the elementary fields should carry a
vector representation, rather than adjoint, in order that the composite
singlet currents reproduce the $hs(4)$ spectrum (\ref{minbHS}) and the
concrete 
proposal is that the boundary theory is the $O(N)$ bosonic vector model
in $d=3$. This essentially means that one identifies the Planck length
in AdS$_4$ with $1/\sqrt{N}$ as $2\kappa_4^2 \sim 1/N$, a normalization
that we will adopt below.

Now, the mass of the conformally coupled scalar in (\ref{confscalar}) is
such that both the {\it regular} and the {\it irregular} boundary modes
could be used to construct a boundary effective action that gives
positive definite 2-pt functions \cite{KW}. Then, one observes that the
conformally coupled scalar  in AdS$_4$ gives the ``spin-zero'' current
of the free O(N) theory only if one uses the {\it non-standard}
dimension $\Delta_-$, which corresponds to the {\it irregular} mode.
Using the {\it standard} dimension $\Delta_+$ appears to give the
large-$N$ interacting fixed point of the model \cite{KP}. The existence
of these two fixed points is related to the fact that the choice of
either $\Delta_-$ or $\Delta_+$ can be imposed by appropriate boundary conditions
corresponding to ``double-trace'' deformations\footnote{With a slight abuse in terminology, 
``double-trace'' operators here mean operators that are quadrilinear in
the elementary $O(N)$ vector fields.} of the boundary theory: the
choice $\Delta_+$ can be imposed by a relevant
``double-trace'' deformation
of the boundary UV fixed point and the choice $\Delta_-$ can
be imposed by an irrelevant deformation of the IR boundary fixed point.
It is interesting to note that the existence of the two different fixed
points in the boundary is also tied to the existence of a large-$N$ limit.\cite{GM,GK}

\section{A fermionic realization of the IR boundary theory and the
  role of parity}

There appears to be another possible construction of the $\Delta_+$
theory above. Suppose we start with the fermionic singleton
$D(1,1/2)$. We have
\begin{equation}
\label{oneonehalf}
\left[D(1,1/2)\otimes D(1,1/2)\right]_A=D(2,0)\oplus \sum_{s=1} D(2s+1,2s)\,.
\end{equation}
This product contains the $D(2,0)$ UIR plus the same tower as in
(\ref{eq:onezerotower}). In this case it seems that the natural
boundary 
theory to associate with this spectrum is a  free $O(N)$ Majorana
fermionic theory. Indeed, at leading order in $1/N$ the dimension of the
basic $O(N)$-singlet $\bar\psi^a\psi^a$ current is one, as is the IR
dimension of the current in the interacting boson model. This
observation appears to suggest that at leading order in $1/N$ the
interacting fixed point of the bosonic $O(N)$ model involves free
fermions.
Moreover, the recent calculation in \cite{tassos1} of the 3-pt functions
of the ``spin-zero'' current in the critical $O(N)$ vector model at
both its UV and IR fixed points for large-$N$ might be viewed as additional
support for such a claim. It was there found, (following earlier work in
\cite{tassos2}),  that the 3-pt function of the ``spin-zero'' current in
the IR fixed point (i.e., the operator with dimension 2) vanishes. This
is consistent with the fact that the 3-pt function of the current
$\bar{\psi}^a\psi^a$ is zero in the free fermionic theory.
Furthermore, one may consider the ``double-trace" deformation
$(\bar\psi\psi)^2$ of the free fermionic boundary theory. From an RG
point of view this is an irrelevant deformation and therefore it is
consistent with the fact that the free fermionic theory corresponds to an
IR fixed point \cite{Zinn-Justin}. We can then ask what is the UV fixed
point at the other end of this irrelevant deformation? The answer is that
such a UV fixed point involves, at leading-$N$, exactly the spectrum
(\ref{eq:onezerotower}) and therefore seems to correspond to the free
$O(N)$ vector model.

The above picture is appealing but overlooks the
subtle role of parity.\footnote{We are indebted to P. Sundell for
extensive discussions that led to the clarification of the role of
parity.} First, let us note that the Legendre transform in AdS/CFT is simply implementing
the ``intertwining'' operation --- i.e., interchanging representations of
dimensions $\Delta$ with those of dimension $d-\Delta$ \cite{Dobrev}. This is
related to a conformal inversion of the form 
$x^\mu\to x^\mu/x^2$ which explains the UV--IR relationship. However, there is an
  additional discrete parity transformation (i.e., reflection in one
  of the spatial coordinates), necessary to bring the
 inversion  into $SO(3,2)$ \cite{Koller}. 
Starting with a bulk scalar as in (\ref{confscalar}) the two
boundary theories corresponding to the choices $\Delta_-$ and
$\Delta_+$ are related by a Legendre transform \cite{KW} and hence the
two different boundary UIRs $D(1,0)$ and $D(2,0)$ are both scalars
i.e. they have positive parity. The crucial point is now that a 
free-field representation of (\ref{eq:onezerotower}) with elementary
scalars is only
possible when $D(1,0)$ has positive parity, while a free-field
fermionic representation of (\ref{oneonehalf}) is possible when
$D(2,0)$ has negative parity. Therefore, the fixed
points corresponding to the Legendre transforms of the free bosonic
and free fermionic theories do not appear to correspond to free field theories, even for
large-$N$. It is intriguing, however, that the 3-pt function of the
parity-even UIR $D(2,0)$ at the IR point of the $O(N)$ vector model
vanishes.

\section{${\cal N}=1$ Higher-Spin Theory in AdS$_4$ and
  the ${\cal N}=1$ $O(N)$ Vector Model in $d=3$}

Our aim here is to discuss the holography of the ${\cal N}=1$ supersymmetric
higher-spin theory and show that the proposal of \cite{KP} and its
intriguing properties arise as special cases. Supersymmetric versions of
higher spin theories have been recently constructed \cite{HSrecent}. The ${\cal
  N}=1$ theory $hs(1,4)$ is built from the two 
singleton UIR's $D(1/2,0)$ and $D(1,1/2)$  and
its spectrum is given by 
\begin{eqnarray}
\left[D(1/2,0)\otimes D(1/2,0)\right]_S&=&D(1,0)\oplus \sum_{s=1} D(2s+1,2s)\,,\\
\left[D(1,1/2)\otimes D(1,1/2)\right]_A&=&D(2,0)\oplus \sum_{s=1} D(2s+1,2s)\,,\\
D(1,1/2)\otimes D(1/2,0)&=&D(3/2,1/2)\oplus \sum_{s=0} D(5/2+s,3/2+s)\,.
\end{eqnarray}
Given the successes of the bosonic theory, it is not hard to suggest that

\begin{quote}
The minimal ${\cal N}=1$
higher-spin theory on $AdS_4$ is dual to the singlet part of the ${\cal N}=1$
supersymmetric $O(N)$ vector model in $d=3$.
\end{quote}

The spectrum of the ${\cal N}=1$ minimal higher-spin theory is
arranged into massless $Osp(1|4)$ supermultiplets
\cite{Heidenreich}. There is one massless\footnote{Massless refers
to the fermion; the bosons are conformally coupled with $m^2=-2$.}
``Wess-Zumino'' multiplet that contains {\it both} scalar UIR's
$D(1,0)$ and $D(2,0)$. In the $AdS_4$ realization of the theory
these correspond to two conformally coupled scalars $h^{(+)}_1$ and
$h_2^{(-)}$. The supermultiplet is completed by a bulk fermion field
$\Psi$, the $D(3/2,1/2)$ UIR. One can easily write the quadratic
part of the action for that multiplet as \cite{MV} (see also \cite{FS}
for a related discussion) 
\be
\label{bulkaction} I_4
=N\int\rmd^4x\sqrt{-g}\left(-\frac{1}{2}(\partial h_1^{(+)})^2 +
(h_1^{(+)})^2 -\frac{1}{2}(\partial h_2^{(-)})^2 +
(h_2^{(-)})^2+\frac12\bar{\Psi}\slash\!\!\!\!{D}\Psi +\cdots\right)\,. 
\ee
The dots on the right hand side of (\ref{bulkaction})
corresponds to the kinetic terms of higher-spin fields as well as
interactions. These are computable, at least in principle. The
scalars $h^{(+)}_1$ and $h^{(-)}_2$ are real while the fermion $\Psi$ is
Majorana. Our notation is explained fully in the Appendix.

Notice now that, as also explained in the Appendix, the UIR's
$D(1,0)$ and $D(2,0)$ in (\ref{bulkaction}) have opposite parity.\cite{BF} This is not explicit in the bulk action (\ref{bulkaction}), but is implied by supersymmetry (and consequently by boundary conditions at $r\to 0$ necessary to preserve supersymmetry). By parity (which could also be referred to as a discrete chiral symmetry), we mean a discrete element which we take to be $(x^0,x^1,x^2,x^3)\to (x^0,x^1,-x^2,x^3)$. Without loss of
generality, we choose hereafter to assign positive parity to
$h_1^{(+)}$ and negative parity to $h_2^{(-)}$ as indicated
by the superscripts. Moreover, the two two-component
Majorana spinors inside $\Psi$ transform under parity with
opposite signs. 

We see then that the choice of boundary
conditions determines essentially the nature of the boundary theories
dual to (\ref{bulkaction}). One of the two possible duals is the
free ${\cal N}=1$ $O(N)$ vector model  in $d=3$ whose action in
superspace is given by (\ref{N1action}). In this case, the bulk
higher-spin currents in $AdS_4$ correspond to $O(N)$-singlet
bilinears of the elementary boundary superfield $\Phi^a$ that may be
represented as $\Phi^a D_{(i_1}\ldots D_{i_{4n})}\Phi^a$, where
the on-shell real superfield is given by 
\be \label{Phionshell}
\Phi^a(x,\theta) = \varphi^a(x) -\bar{\theta}\psi^a(x)\,,\,\,\,
a=1,2,...,N\,.
\ee 
In particular, the terms in the bulk action
depicted in (\ref{bulkaction}) should reproduce holographically
the generating functional for correlation functions of the
``spin-zero'' current $J$, which is, on-shell 
\be 
\label{Phi2} J(x,\theta)=\frac12\Phi^2(x,\theta)
=\frac12\left(\varphi^a\varphi^a\right)(x)
-\bar{\theta}\left(\psi^a\varphi^a\right)(x)
-\frac12\bar{\theta}\theta \left(\bar{\psi}^a\psi^a\right)(x)\,. 
\ee

In the next section, we consider further the role of boundary conditions in the holography of 
the ${\cal N}=1$ $O(N)$ theories.

\section{Boundary Conditions and Deformations: General Remarks}
\label{sec:bdycond1}

The ${\cal N}=1$ supersymmetric $O(N)$ vector model in the boundary
has been studied extensively in the 
literature using mainly a $\sigma$-model approach \cite{Gracey}. It
possesses two critical points at large-$N$, one free and one interacting, which have
various supersymmetric and non-supersymmetric
deformations that could be considered. We consider several such deformations in Section \ref{sec:bdycond2}, although the 
``double-trace'' deformations of the supersymmetric fixed points are of particular interest.
From the bulk point of view,
these correspond to suitable choices of the boundary conditions for
scalars and spinors. We review this  in the present context below\cite{witten,BSS}.

\subsection{Scalars}

We consider the conformally coupled scalars in (\ref{bulkaction}) whose
asymptotic behavior near the AdS$_4$ boundary at $r\rightarrow 0$ is
\be
\label{hks}
h_k^{(\pm)}=r\a_k^{(\pm)}(x)+r^2\b_k^{(\pm)}(x)+\ldots\,,\,\,\,\,\,k=1,2\,.
\ee
Requiring that the bulk solutions vanish as $r\rightarrow\infty$ one obtains
\begin{eqnarray}
\label{b1a1}
&& \a_1^{(+)}(x)=-\frac{1}{2\pi^2}\int d^3y\ \frac{1}{(x-y)^{2}} \b^{(+)}_1(y)\,,\\
\label{b2a2}
&& \b^{(-)}_2(x)=-\frac{1}{2\pi^2}\int d^3y\ \frac{1}{(x-y)^{4}} \a_2^{(-)}(y)\,.
\end{eqnarray}
Therefore, each bulk solution (\ref{hks}) depends on one arbitrary
function and when substituted back into the bulk action yields a well
defined boundary functional. Usually one is forced to consider a
functional of the $\a$'s and this is referred to as choosing the {\it
regular} boundary conditions. However, as pointed out in \cite{KW} there
are cases where a functional of the $\b$'s is perfectly acceptable, and
this is referred to as choosing the {\it irregular} boundary conditions.
Then, the functionals of the $\a$'s and the $\b$'s are related by a
Legendre transform as is implied by (\ref{b1a1}) and (\ref{b2a2}),
which preserves the parity assignments.
Furthermore, choosing as boundary sources one of the $\a$'s or the
$\b$'s, the other becomes the expectation value of the boundary
operator. Explicitly, to get the free boundary fixed point we want
here to choose $\b^{(+)}_1(x)$ as a  source for
the boundary operator $J^{(+)}_1(x)$, and then $\a^{(-)}_1(x)$ represents the
one-point function $\langle J^{(-)}_1(x)\rangle$. Conversely, for $J^{(-)}_2(x)$ it
is  $\a^{(-)}_2(x)$ that sources the field, and $\b^{(-)}_2(x)$ gives the
corresponding one-point function. Thus the supersymmetric boundary
condition on the bulk scalar fields of the action (\ref{bulkaction}) should be
\be
\label{susybc}
\a^{(+)}_1(x)=\b^{(-)}_2(x)=0\,.
\ee

Now, if we want to perturb the Lagrangian of the boundary theory, we
take a suitable boundary condition for $h^{(\pm)}_k$. The precise form of the
boundary condition will determine the actual perturbation. One simple
deformation is the ``single trace" $\int d^3x f^{(+)}_1(x)J^{(+)}_1(x)$, whereby we
simply set the value of $\b^{(+)}_1=f^{(+)}_1$. Similarly, if we want the single
trace deformation $\int d^3x f^{(-)}_2(x)J^{(-)}_2(x)$, the appropriate boundary
condition is $\a^{(-)}_2=f^{(-)}_2$. For more general deformations the prescription
given by Witten\cite{witten} may be summarized as follows. If we wish
to generate the perturbation
\begin{equation}
\label{Wpert}
\int d^3x\ W(x; J^{(+)}_1,dJ^{(+)}_1,\ldots,J^{(-)}_2,dJ^{(-)}_2,\ldots)\,,
\end{equation}
we impose the boundary conditions
\bea
\label{bcW}
\b^{(+)}_1&=&\frac{\delta W(x;
  \alpha^{(+)}_1,d\alpha^{(+)}_1,\ldots,\beta^{(-)}_2,d\beta^{(-)}_2,\ldots)}{\delta
  \alpha^{(+)}_1}\,,\\
 \a^{(-)}_2&=&\frac{\delta W(x;
  \alpha^{(+)}_1,d\alpha^{(+)}_1,\ldots,\beta^{(-)}_2,d\beta^{(-)}_2,\ldots)}{\delta
  \b^{(-)}_2}\,.
\eea

For completeness, let us give an example of how this result comes about. Consider for
simplicity a single bosonic field $h$ conjugate to the boundary
operator ${\cal O}$ with dimension $1/2<\Delta<3/2$. The field $h$ has the behavior
\be
\label{hsimple}
h\sim r^{3-\Delta} \alpha_0(x)+r^{\Delta}\beta_0(x)+\ldots\,,
\ee
while $\a_0$ and $\beta_0$ are related as
\be
\label{expvalue}
\beta_0(x)\equiv\langle {\cal O}\rangle_{\a,0} \sim\int d^3y\
\frac{1}{(x-y)^{2(3-\Delta)}} \alpha_0(y)\,.
\ee
Add a boundary interaction $\frac{f}{2N}\int d^3y\ {\cal O}(y)^2$
 and consider the calculation of
\be
\langle {\cal O}(x)\rangle_{\a,f}=\langle {\cal O}(x)
e^{i\frac{f}{2N}\int d^3y\ {\cal O}(y)^2}\rangle_{\a,0}\,.
\ee
We can proceed by expanding the exponential. The crucial point
here is the assumption of a large-$N$ expansion
such that only the leading term in the OPE ${\cal O}(x) {\cal
  O}(y)\sim \frac{N}{(x-y)^{2\Delta}} I+\ldots$ contributes in the
large-$N$ limit. Then we
derive
\bea
\langle {\cal O}(x)\rangle_{\a,f}&=&\langle {\cal O}(x)\rangle_{\a,0}+f
\int d^3y\ \frac{1}{(x-y)^{2\Delta}}\langle {\cal O}(y)
\rangle_{\a,0}\nonumber \\
\label{Oalpha}
&& +f^2\int d^3y\ \frac{1}{(x-y)^{2\Delta}}\int d^3z\
\frac{1}{(y-z)^{2\Delta}}\langle {\cal O}(z) \rangle_{\a,0}+\ldots\,.
\eea
Now, we can multiply (\ref{Oalpha}) by the inverse kernel of
(\ref{expvalue}) to obtain
\be
\int d^3x\ \frac{1}{(y-x)^{2(3-\Delta)}}\langle {\cal O}(x)\rangle_{\a,f} = \alpha_f(y)\,,
\ee
and we then have
\be
\alpha_f(y)=\alpha_0(y)+f\beta_0(y)+f^2\int d^3x\
\frac{1}{(y-x)^{2\Delta}} \beta_0(x)+\ldots\,, 
\ee
and thus
\be
\label{afbf}
\alpha_f=\alpha_0+f\beta_f\,.
\ee
Setting then the sources to zero in the unperturbed theory
$\alpha_0=0$, we arrive at the advertised boundary condition (\ref{bcW}). It is
clear that this derivation will generalize to an arbitrary
functional $W$.

The derivation above elucidates also the meaning of choosing the {\it
  irregular} boundary conditions, i.e., considering a functional of $\b_0$ in the case of
  (\ref{hsimple}). Formally, this corresponds to considering
  the Legendre transform of the standard functional of $\a_0$ that is
  the generating functional of operators
  $\hat{\cal O}$ with dimension $3-\Delta$. Consider now a boundary
  interaction $\frac{g}{2N}\int d^3y\ \hat{\cal O}(y)^2$. Following
  the same reasoning as above we find
\be
\label{agbg}
\b_g =\b_0 +g\a_g\,.
\ee
We now see that $f\rightarrow \infty$  in the {\it regular} choice
  (\ref{afbf}) leads to the unperturbed boundary condition $g=0$ in
  (\ref{agbg}) and hence to the {\it irregular} choice. The reverse
  also holds true and $g\rightarrow \infty$ in the {\it irregular}
  choice leads to the unperturbed $f=0$ {\it regular} choice (\ref{afbf}).

\subsection{Spinors}\label{sec:spinorbc}

Next we consider a massless bulk spinor 
 $\Psi$.
The Dirac equation has the form
\begin{equation}
\left[ \Gamma^3 (r\partial_r-3/2)+r\Gamma^\mu\partial_\mu\right]\Psi=0\,.
\end{equation}
This can be transformed into a second order equation whose general
on-shell solution is
\be
\label{Psisolut}
\Psi(x,r)=\int \frac{d^3p}{(2\pi)^3}\ e^{ip\cdot x} r^2 K_{1/2}(pr)
\pmatrix{ a_+(p)\cr a_-(p)}\,,
\ee
where $a_\pm$ are two-component spinors.
Substituting back into the first order equations, we find
asymptotically\footnote{The precise form is
  $\frac{(yD+1/2)K_{1/2}(y)}{K_{1/2}(y)} a_\mp=\mp
  iy\underline\gamma^\mu n_\mu a_\pm
$.}
\be
a_-= i \underline {\slash\!\!\!{n}} a_+\,,\,\,\,\,\, n_\mu=\frac{p_\mu}{p}\,.
\ee
The asymptotic behavior of (\ref{Psisolut}) near the boundary is
\be
\Psi(x,r)\sim r^{3/2}\pmatrix{u_+(x)\cr u_-(x)}\,,
\ee
where
\bea
\label{upm}
&&u_-(x)=+\int d^3y\ \underline{G}(x-y)\ u_+(y)\,,\\
\label{Gz}
&&\underline{G}(z)=-\underline{G}(z)^{-1}=i\int \frac{d^3p}{(2\pi)^3}\
e^{ip\cdot z}\frac{\underline{\slash\!\!\!{p}}}{p}
=-\frac{1}{\pi^2}\frac{\slash\!\!\!x}{x^4}\,. 
\eea
Notice now that under the parity transformation of (\ref{parity1}),
$u_+$ and $u_-$ transform with different signs, while the kernel
$(\ref{Gz})$ remains invariant. The latter is actually proportional to
the 2-pt function of the boundary fermions, and is sometimes referred
to as an intertwiner \cite{Dobrev}.

Since the boundary action is of the form  \cite{Henneaux}
\be
\label{bferm}
\int d^3x \left(\bar{u}_+ u_+ - \bar{u}_- u_-\right)\,,
\ee
it appears that which of $u_\pm$ one calls $\langle {\cal
  O}_{3/2}\rangle$ and which one calls the source is a matter of
choice. However, the supersymmetry structure indicates the proper
interpretation. Recall that in the bulk, the supercharge splits as in (\ref{eq:Qsplit})
\begin{equation}
Q=\pmatrix{q\cr s}\,,
\end{equation}
where $q$ is the supercharge and $s$ is the superconformal
generator. After making a choice of parity assignments for
the bulk scalars, e.g., assign negative parity to $h^{(-)}_2$, the
supersymmetry in the bulk requires that $q h^{(-)}_2\sim
\Psi$. (in the notation of the Appendix, $h^{(-)}_2\sim j_2$). Now, we
remember that to get the free $O(N)$ theory in the boundary, we use the
{\it regular} boundary conditions for $h^{(-)}_2$. Therefore in this case we
should identify $u_+$
with the vev and $u_-$ with the source. The opposite will hold true
when we want to find the strongly-coupled $O(N)$ boundary theory.

Consider then the classically marginal ``double-trace" operator
\begin{equation}\label{eq:fermdbl}
e^{i\frac{E}{2N}\int d^3x\  \bar{\cal O}_{3/2}{\cal O}_{3/2}}\,.
\end{equation}
To determine the boundary condition to which it corresponds 
we follow the arguments in the previous subsection and write
\be
u_-^{(E)}(x)=u_-^{(0)}(x)+iE\int d^3y\ iG(x-y)u_-^{(0)}(y)+\ldots\,.
\ee
Assuming then an OPE of the form
\be
{\cal O}_{3/2}^{i}(x){\cal O}_{3/2}^{j}(y)\sim NiG^{ij}(x-y) I +\ldots\,,
\ee
and a large-$N$ expansion we arrive at the condition
\be
\label{uE}
u_-^{(E)}=u_-^{(0)}+Eu_+^{(E)}.
\ee
Note now that (\ref{uE}) is a boundary condition that does not
preserve parity,
but $E\to\infty$ clearly corresponds to switching $q$ and $s$, and thus we
would expect a rearrangement of the supermultiplet, if the theory is supersymmetric at $E\to\infty$.

\section{Boundary Conditions and Deformations: ${\cal N}=1$
  Higher-Spin Theory}\label{sec:bdycond2}

Now we are ready to present our concrete proposal by considering various
boundary conditions  for the bulk fields that appear in the action
(\ref{bulkaction}). First we will consider the large-$N$ duals of
the bulk theory which means that we discuss only the tree level
bulk action. In section \ref{sec:1overN}, we will discuss the effects of bulk loops on the
higher-spin gauge symmetry.

\subsection{Large-$N$ duals}

\subsubsection{The free ${\cal N}=1$ Supersymmetric $O(N)$ vector model}

To obtain the free ${\cal N}=1$ Supersymmetric $O(N)$ vector model
in the boundary we consider the {\it irregular} boundary
condition for $h_1^{(+)}$ and the {\it regular} one for $h_2^{(-)}$.
Making also an appropriate choice for the boundary condition of
the bulk fermion, as explained in Section \ref{sec:spinorbc}, we preserve
supersymmetry. Moreover, as we will argue in Section \ref{sec:1overN} these
boundary conditions will not break the higher-spin gauge symmetry even
after bulk loops are taken into account.

\subsubsection{The strongly coupled ${\cal N}=1$ Supersymmetric
$O(N)$ vector model}\label{sec:strcplfp}

In this case, we use the {\it regular} boundary condition
for $h_1^{(+)}$ and the {\it irregular} one for $h_2^{(-)}$. Then,
with the opposite choice from above for the boundary conditions of
the bulk fermions we still preserve supersymmetry. However, now
the boundary ``spin-zero" current multiplet cannot be represented
by free fields due to the specific parity assignments. This ${\cal N}=1$ theory corresponds
to the large-$N$ limit of the IR fixed point of the $O(N)$
vector model. The reason is that, as we will argue in Section \ref{sec:1overN}, the
specific boundary conditions chosen here will break the higher-spin
gauge symmetry for subleading-$N$ when bulk loops are taken into
account. In this way the currents in the boundary theory will acquire
anomalous dimensions of order $1/N$. 

\subsection{Deformations}

Next, we consider deformations of the boundary Lagrangian.

\subsubsection{Mass deformations}

The simplest deformation of the free ${\cal N}=1$ theory that one
could consider is a boundary condition corresponding to adding a mass
term for elementary fields in the boundary theory. For example, a
boundary fermion mass term will clearly lead to an infrared theory
containing only currents built out of the elementary bosons $\varphi^a$, once $1/N$ corrections are taken into account. From the
bulk point of view, this physics should be reproduced classically, by a suitable ``domain wall''.  In particular, one must be able to see
that all higher spin bulk fields that were coupled to boundary operators
involving fermions (i.e., $h^{(-)}_2$ and $\Psi$ and their higher-spin
partners) are made massive by the choice of boundary condition. In
this way, one is left with the spectrum of the $hs(4)$ bosonic higher
spin theory. Similarly, a boundary boson mass term should leave only
the fermionic currents in the boundary theory, or in the bulk, only
$h^{(-)}_2$ and its higher-spin partners. 

\subsubsection{A marginal double-trace perturbation}

Given the established connection between the Legendre transformation and double trace operators, it is clearly of interest to study these in the present context. There are two distinct choices that we will identify. 

First, a (classically) marginal deformation of the free boundary theory is 
 \begin{equation}
 \label{marginal}
 \int d^3x\ \left\{ g_3 J_1^{(+)}J_2^{(-)}+g_4 \bar{J}_{3/2}^{(\pm)}J_{3/2}^{(\pm)}\right\}\,.
 \end{equation}
  This is supersymmetric along the line $g_3= -2g_4$ in which case it corresponds to the
  deformation $\int d^3x \,d^2\theta\, J^2$. It is easily seen
  that this deformation violates parity \cite{KM}. It is interesting
  to ask where the deformation (\ref{marginal}) leads the free
  $O(N)$ vector model to, for large values of the couplings $g_3$,
  $g_4$ and large $N$. One way to answer
  this is to recall that
  (\ref{marginal}) can actually be imposed via an appropriate
  boundary condition on the bulk fields. In the notation of the
previous section this is
  \begin{equation}
  \alpha^{(-)}_2=g_3 \alpha^{(+)}_1,\ \ \ \ \
  \b^{(+)}_1=g_3\b^{(-)}_2, \ \ \ \ \ \ u_-=-g_4 u_+ \,.
  \end{equation}
We now see that in the supersymmetric case the limit
of large coupling constant and large $N$ just leads to the strongly coupled
theory of Subsection \ref{sec:strcplfp}. Thus, even though the deformation
(\ref{marginal}) breaks parity, at large $N$ we end up with an
${\cal N}=1$ supersymmetric theory.

It is also possible that the deformation (\ref{marginal}) is actually {\it exactly} marginal. This sort of possibility
was mentioned in Ref. \cite{witten} in the bosonic theory, where it
was noted that AdS$_4$ apparently remains a solution for any value of
the coupling $\int d^3x {\cal O}_{\Delta_-}{\cal O}_{\Delta_+}$. In
the present case, the situation is even better: eq.   (\ref{marginal})
is a deformation of a {\it free} CFT and thus marginality can be
investigated perturbatively. We are not aware of literature related
directly to this question, but it should not be too difficult an issue
to settle. 

\subsubsection{A supersymmetry breaking deformation}

Another interesting ``double-trace'' deformation of the free
theory is
\begin{equation}
\label{susybreak}
\frac12\int d^3x\ \left\{
g_1(J_1^{(+)})^2+g_2(J_2^{(-)})^2+g_3\bar{J}^{(\pm)}_{3/2}J^{(\pm)}_{3/2}\,.
\right\}
\end{equation}
This is a closer analogue to the double trace deformation considered for the bosonic theory.
It corresponds to the boundary conditions 
\be 
\label{bcsbr}
\alpha^{(-)}_2=g_2\beta^{(-)}_2\,,\,\,\,\,\, \b^{(+)}_1=g_1\a^{(+)}_1\,,\,\,\,\,\,
u_-=-g_3 u_+ \,,
\ee which of course break supersymmetry. $g_1$
corresponds to a relevant deformation of the free $O(N)$ fixed
point while the $g_2$ deformation looks irrelevant from an RG
point of view. Again, we can ask where the deformation
(\ref{susybreak}) leads the free boundary theory to, for large
values of the couplings and large $N$. This can be answered using information from the
bulk. Namely, we see from the boundary conditions (\ref{bcsbr})
that for large values of the coupling constants one is apparently led to the
strongly coupled theory of Subsection \ref{sec:strcplfp} again! This time,
the boundary conditions for the scalars are parity preserving
while the one for the fermion is parity non-preserving. The
remarkable result is that despite the fact the the deformation
(\ref{susybreak}) breaks supersymmetry, at large $N$ we recover an
${\cal N}=1$ supersymmetric theory,  at its strongly
coupled fixed point. This is perhaps not as surprising as it seems, as
the RG interpretation of the deformation (\ref{susybreak}) is rather
unusual. Indeed, due
to the structure of the Wess-Zumino multiplet, one may view the free boundary
theory as being ``half at the UV fixed point" (the part involving the
bosons) while the other half is at its ``IR fixed point" (the part
involving the fermions). 

\subsection{Subleading-$N$ and the breaking of higher-spin gauge
symmetry}\label{sec:1overN}

In the previous subsection we argued that the tree level ${\cal
N}=1$ higher-spin theory on AdS$_4$ leads to {\it two} boundary
3d CFTs whose composite operators have {\it exactly} the same
spectrum of dimensions. The only property that distinguishes the
two boundary theories, at large $N$,  is the parity assignment of the operators in
the supermultiplets. One choice of parity assignments leads to the
free $O(N)$ vector model while the other choice leads to a
strongly coupled version of the $O(N)$ vector model. The
generating functionals of the two theories are related by a
Legendre transform.

The distinction between the two
theories should become more evident when considering bulk loop
corrections or, equivalenty corrections subleading in $1/N$ in the boundary theory. 
In other words, we expect that starting with the boundary action
(\ref{bulkaction}) and considering the boundary conditions that
lead to the free boundary theory, bulk loops do 
not break the higher-spin gauge symmetry. That is,  the bulk
higher-spin fields remain massless while the corresponding boundary
currents remain conserved. On the other hand, considering the
boundary conditions that lead to the strongly coupled boundary
theory we expect that the bulk loops will render the higher-spin
fields massive (Higgsing) and the corresponding boundary currents
non-conserved \cite{Ruhl}.

The mechanism by which the phenomenon described above takes place
is a generalization of the mechanism discussed in \cite{GPZ} for
the case of the minimal bosonic higher-spin theory on AdS$_4$. The basic physics arises from the fact that a representation that is massive from the bulk point of view satisfies
\begin{equation}
D(\Delta,s)\rightarrow  D(s+1,s)\oplus D(s+2,s-1)
\end{equation}
as $\Delta\to s+1$. This can be taken to imply that in order for a spin $s$ field to become massive, there must be a suitable Goldstone mode transforming as $D(s+2,s-1)$, it must be of the correct parity, and there must be a suitable coupling present in the bulk effective action.

Here we elucidate this argument further by discussing it from the boundary
point of view. Conformal invariance requires that a boundary spin $s$
current $J^{\m_1,..,\m_s}$ with dimension $\Delta_s=s+1$ is
also conserved. Non-conservation of this current appears in the
form of an anomalous dimension, $\Delta_s \rightarrow s+1+\gamma$. Let us
assume that $\gamma\sim O(1/N)$. Then, the
non-conservation of a boundary current means that there is an operator equation of the form
\be \label{nonconserv}
\partial_{\m_1} J^{\m_1,..,\m_s}(x)\sim \frac{1}{\sqrt{N}}
T^{\m_2,..,\m_s}(x)\,. 
\ee 
The current $T^{\m_2,..,\m_s}(x)$ has dimension $s+2$ to
leading order in $1/N$. For non-coincident points, 
(\ref{nonconserv}) leads to 
\be 
\label{anselmi}
\langle\partial_{\m_1} J^{\m_1,..,\m_s}(x_1)\,\partial_{\n_1}
J^{\n_1,..,\n_s}(x_2)\rangle \sim \frac{1}{N} \langle
T^{\m_2,..,\m_s}(x_1)T^{\n_2,..,\n_s}(x_2)\rangle\,.
 \ee 
Then, conformal invariance determines\cite{Anselmi} the form of both sides in
(\ref{anselmi}) and a tree-level calculation of the rhs of (\ref{anselmi}) yields the
$1/N$ result for $\gamma$. What is crucial for us is that 
equations such as (\ref{nonconserv}) can exist in the theory
only if a current $T^{\m_2,..,\m_s}(x)$ with the appropriate
dimension $s+2$ and parity can be constructed.

Now let us consider the ${\cal N}=1$ higher-spin theory on AdS$_4$.
In the bulk effective action, there might be terms of the
schematic form\footnote{These couplings should be suitably supersymmetrized, but we will
not consider the details of this here.}
\begin{equation}
\label{bulkcubic}
\frac{1}{\sqrt{N}}W^{a_1,..,a_s}W_{a_3,..,a_s}\partial_{a_1}\partial_{a_2} h\,,
\end{equation}
where $W$ are higher-spin currents and $h$ is either of the two
conformally coupled bulk scalars. If such a term exists, then  it can give rise to a boundary 3-pt function
of the form \be 
\label{3pthiggs} \langle
\partial_{\m_1}J^{\m_1,...,\m_s}(x_1)
J_{\m_3,...,\m_s}(x_2)\partial_{\m_2}J(x_3)\rangle\,.
\ee 
This 3-pt
function will be non-zero and would correspond to 
(\ref{anselmi}) if in the OPE of $J_{\m_3,..,\m_s}$ with $J$ there
exists an operator such that its derivative produces the current
$T^{\m_2,..,\m_s}(x)$. Let us study this in more detail. The OPE
in question is of the form
\be \label{ope} J_{\m_3,...,\m_s}(x_2) J(x_3)
\approx {\cal T}_{\m_3,..,\m_s}(x_3) +(x_{23})^\n {\cal
S}_{\n\m_3,..\m_s}(x_3)+ ...\,,
 \ee 
where the dots stand for higher
descendants and other operators.
Now we have to take into account the dimension of $J(x)$. When
$J(x)$ has dimension 2 the operator ${\cal T}$ in
(\ref{ope}) has dimension $s+1$ and its derivative has dimension
$s+2$ and could be a candidate for $T$. But only when $J$ has
matching parity can this OPE produce the correct $T$. On
the other hand, when $J$ has dimension 1, it is the operator ${\cal
  S}$ in  (\ref{ope}) that has dimension $s+1$ and 
therefore its derivative might give rise to $T$. In this case
however, only when $J$ has opposite parity can the correct
$T$ arise. Moreover, it is easy to see that when $s=2$ the OPE
(\ref{ope}) is between the same boundary scalar. Therefore, whichever boundary
condition one chooses, the correct $T$ can never arise. Therefore, the
boundary energy momentum tensor remains always conserved  as it
should.

\section{Summary and Discussion}

We have presented a concrete proposal for the holographic dual of the
${\cal N}=1$ higher-spin gauge theory on AdS$_4$. We have argued that the
boundary theory is the ${\cal N}=1$ supersymmetric $O(N)$ vector model
in three dimensions. Both regular and irregular bulk modes are
necessary for this holography and their interplay unveils interesting
phenomena. In particular, the unique bulk theory gives rise to two
boundary theories that are the free and interacting fixed points of
the $O(N)$ vector model. At large-$N$, the boundary theories are
distinguished only by the parity assignments in the
supermultiplets. For subleading-$N$, only the boundary conditions that
give the interacting $O(N)$ vector model in the boundary will Higgs the
massless higher-spins and give rise to a boundary theory where all
higher-spin currents acquire anomalous dimensions. 

We studied various boundary conditions that correspond to
``double-trace'' deformations of the free boundary
theory. Particularly intriguing is the fact that supersymmetry
breaking boundary deformations lead for large-$N$ to a supersymmetric
theory. This phenomenon is tied to the fact that the free boundary
theory may be viewed as being ``half in a UV fixed point and half
in an IR fixed point.'' We expect that a similar phenomenon occurs
in the case of the the ${\cal N}=1$ SCFT in four dimensions obtained
holographically from the compactification of IIB SUGRA on
AdS$_5\times$T$^{1,1}$ \cite{KW2}. We have also noted the possible existence of a line of fixed points in this model.

One would like to think that some of the salient features of this
special holography are connected with the higher-spin gauge
symmetry. In particular the fact that this holography gives rise to a
free boundary theory is presumably a feature of higher-spin
theories. Nevertheless, it appears that once we have information
about the free boundary theory we also have information about an
interacting boundary theory. This follows from the fact that the
higher-spin multiplet includes simultaneously ``shadow'' UIRs of the
conformal group and therefore describes at the same time UV and IR
properties of the boundary theory. It would be interesting to study
further our proposal and discuss supermultiplets containing currents
with higher spins, in particular the energy momentum tensor. It is
also be of interest to study the thermodynamics and the $O(N)$
symmetry breaking pattern of the boundary theory from the bulk point
of view.

\begin{acknowledgments}
We thank P. Sundell, M. Vasiliev and S. Ferrara for useful discussions. The work of RGL is supported in part by U.S. Department of Energy grant DE-FG02-91ER40677.
\end{acknowledgments}

\begin{appendix}
\section*{Appendix}
\section{Conventions}

There is a real basis of generators for the $(-++)$ Clifford  algebra
\begin{equation}
\label{eq:Clif}
\underline{\gamma}^\mu=\{ i\sigma_2,\sigma_1,\sigma_3\}\,.
\end{equation}
Because of the reality of the generators of the Clifford algebra, we
can take Majorana spinors satisfying
\begin{equation}
\label{ }
\psi=\hat{c}\bar\psi^T\,.
\end{equation}
We can take $\hat{c}=\underline{\gamma}^0\equiv \epsilon^{\a\b}$, and thus the Majorana
condition just reduces to
$\psi^*=\psi$.
Because of
the choice of basis for the Clifford algebra, we have equations
like
\be
\bar\psi\eta=
\psi^T\gamma^0\eta=\psi_\a\epsilon^{\a\b}\eta_\b=-\epsilon^{\b\a}\psi_\a\eta_\b
=-\psi\eta\,.
\ee
Using real Poincar\'e generators, the ${\cal N}=1$ supersymmetry
generators can be taken to satisfy
\be
\label{qN1}
\left\{ q_\alpha,q_\beta\right\} = 2(\underline\gamma^\mu
\underline{\hat c}^{-1})_{\alpha\beta} P_\mu\,.
\ee
A superspace representation for these generators is
\be
q_\alpha=-(\hat c^{-1})_{\alpha\beta}
\frac{\delta}{\delta\theta_\beta}
+(\underline\gamma^\mu\theta)_\alpha\partial_\mu\,.
\ee
The general ${\cal N}=1$ superfield is written
\be
\label{Phi}
\Phi(x,\theta)=\varphi(x)-\bar\theta\psi(x)+\frac12\bar\theta\theta F(x)
\ee
and the supersymmetry transformations are
\be
\delta\varphi = \bar\epsilon\psi\,,\,\,\,
\delta\psi = F\epsilon+\partial_\mu\varphi\underline\gamma^\mu\epsilon\,,\,\,\,
\delta F= -\bar\epsilon\underline\gamma^\mu\partial_\mu\psi\,.
\ee
The ``spin-zero'' current is defined as
\be
\label{Jcurrent}
J=\frac12\Phi^2=\frac12(\varphi\varphi)-\bar\theta \left[
  (\varphi\psi)\right] +\frac12\bar\theta\theta \left[ (\varphi
  F)-\frac12(\bar\psi\psi)\right]\,,
\ee 
therefore on-shell we have 
\be \label{Js}
J_1^{(+)}=\frac12(\varphi\varphi)\,,\,\,\,\,
J_{3/2}^{(\pm)}=(\varphi\psi)\,,\,\,\,\, J_2^{(-)}=-\frac12
(\bar\psi\psi)\,. 
\ee
Then, taking 
\be D_\alpha=(\hat
c^{-1})_{\alpha\beta} \frac{\delta}{\delta\theta_\beta}
+(\underline\gamma^\mu\theta)_\alpha\partial_\mu \,,
\ee 
we find the
free ${\cal N}=1$ Lagrangian as
\be
\label{N1action}
\int d^2\theta \frac12\overline{D\Phi}D\Phi=-\frac12
\left[\eta^{\mu\nu}\partial_\mu\varphi
  \partial_\nu\varphi+\bar\psi\underline\gamma^\mu\partial_\mu\psi+F^2\right]\,.
\ee 
Parity in three dimensions is defined as 
\be \label{parity1}
\hat{P}\psi(x^0,x^1,x^2) =
\eta\left(\Pi\cdot\psi\right)(x^0,x^1,-x^2)\,,\,\,\,\,\eta^2=1 \,,
\ee
and one can verify that a suitable choice is $\Pi \equiv
\gamma^2$. Then one easily finds that the scalar $J_2^{(-)}(x)$ in
(\ref{Js}) is odd under parity while $J_1^{(+)}$ is even, as the
superscripts indicate. In general, ${\cal N}=1$ supersymmetry
requires that the two scalars in the ``spin-zero'' multiplet have
opposite parity. However, only the specific assignments above lead
to a free-field theory representation as in (\ref{Jcurrent}).




Now we would like to extend this to the $AdS_4$ bulk \cite{MV}.
First, we take the AdS$_4$ metric in Poincar\'e coordinates
\begin{equation}
ds^2=\frac{1}{r^2}\left( dr^2+\eta_{\mu\nu} dx^\mu
dx^\nu\right)\,,\,\,\,\,  \m ,\n=0,1,2\,.
\end{equation}
It is simpler to work in $\bR^5$ with metric $g^{AB}=diag(--+++)$,
$A,B=-1,0,1,2,3$ where $Spin(3,2)$ acts
linearly. The most convenient basis is
\begin{equation}
\Gamma^{-1}=\pmatrix{
   0   & -I   \cr
    I  &  0
} \ \ \ \ \ \
\Gamma^\mu=\pmatrix{
  \underline{\gamma}^\mu    & 0   \cr
  0    &  -\underline{\gamma}^\mu
} \ \ \ \ \ \
\Gamma^3=\pmatrix{
 0    & I   \cr
  I    &  0
}\,.
\end{equation}
The $Spin(3,2)$ generators in the spinor representation are
$S^{AB}=\frac{1}{4}\left[ \Gamma^A,\Gamma^B\right]$.
Denoting by $M^{AB}$ the $SO(3,2)$ generators, we can introduce a
supercharge $Q_\alpha$ (here $\a=1,..,4$)
\begin{equation}
\left[ M^{AB}, Q_\alpha\right]=-{(S^{AB})_\alpha}^\beta Q_\beta\,.
\end{equation}
In the given basis, $\hat C=\Gamma^{-1}\Gamma^0$. We have
\begin{equation}
\left\{ Q_\alpha,Q_\beta\right\}=-2(S^{AB}\hat C^{-1})_{\alpha\beta} M_{AB}\,.
\end{equation}
The utility of the chosen basis is that the generator of
$Spin(2,1)\subset Spin(3,2)$  splits
\begin{equation}
S^{\mu\nu}=\pmatrix{ \frac12\underline\gamma^{\mu\nu}&0\cr 
0&\frac12\underline\gamma^{\mu\nu}}\,,
\end{equation}
and thus it is sensible to define
\begin{equation}\label{eq:Qsplit}
Q=\pmatrix{ q\cr s}\,.
\end{equation}
We can then work out the superalgebra relations\footnote{As usual,
  define $D=M_{-1,3}, K_\mu=M_{3,\mu}+M_{-1,\mu},
  P_\mu=M_{3,\mu}-M_{-1,\mu}, L_{\mu\nu}=M_{\mu\nu}$.} (now
$\alpha=1,2$)
\begin{eqnarray}
\left[ L_{\mu\nu}, q_\alpha\right] = -\frac12
(\underline\gamma_{\mu\nu}q)_\alpha & \ \ \ \ & \left[ L_{\mu\nu},
  s_\alpha\right] = -\frac12 (\underline\gamma_{\mu\nu}s)_\alpha\,, \\
\left[ P_{\mu}, q_\alpha\right] = 0 & \ \ \ \ & \left[ P_{\mu},
  s_\alpha\right] = -(\underline\gamma_{\mu}s)_\alpha\,, \\
\left[ K_{\mu}, q_\alpha\right] = +(\underline\gamma_{\mu}q)_\alpha &
\ \ \ \ & \left[ K_{\mu}, s_\alpha\right] = 0 \,,\\
\left[ D, q_\alpha\right] =-\frac12 q_\alpha & \ \ \ \ & \left[ D,
  s_\alpha\right] = +\frac12 s_\alpha\,,
\end{eqnarray}
and
\begin{eqnarray}
\label{qqP}
\left\{ q_\alpha,q_\beta\right\} &=& 2(\underline\gamma^\mu
\underline{\hat c}^{-1})_{\alpha\beta} P_\mu\,,\\
\left\{ s_\alpha,s_\beta\right\} &=& -2(\underline\gamma^\mu
\underline{\hat c}^{-1})_{\alpha\beta} K_\mu\,,\\
\left\{ q_\alpha,s_\beta\right\} &=& 2(\underline{\hat
  c}^{-1})_{\alpha\beta} D - (\underline\gamma^{\mu\nu}
\underline{\hat c}^{-1})_{\alpha\beta}L_{\mu\nu}\,,
\end{eqnarray}
where we have written $\hat C=\pmatrix{0&\underline{\hat c}\cr
  \underline{\hat c}&0}$.
It is clear from the form of the algebra that $Q$ can be taken to be
  Majorana. 
This condition is
\begin{equation}
Q=C{\overline Q}^T\,,
\end{equation}
where $C=\hat C\Gamma^{-1}$, and the condition amounts to $q^*=q, s^*=s$.

The bulk "Wess-Zumino multiplet" has real scalar components $j_0,j_2$
and a Majorana spinor $j_{1\alpha}$. In terms of UIRs of $Osp(1|4)$
this is of the general form \cite{Heidenreich}
\be
\label{Osp14}
D(\Delta,0)\oplus D(\Delta+1/2,1/2)\oplus D(\Delta+1,0)\,.
\ee
Supersymmetry acts on them as
\begin{eqnarray}
q_\alpha j_2&=& {j_1}_\alpha\,,\\
\label{qj1}
q_\alpha {j_1}_\beta &=& (\underline\gamma^\mu\underline{\hat
  c}^{-1})_{\alpha\beta}\partial_\mu j_2 + (\underline{\hat
  c}^{-1})_{\alpha\beta} j_0\,,\\
\label{qj0}
q_\alpha j_0 &=& -(\underline\gamma^\mu\partial_\mu j_1)_\alpha\,.
\end{eqnarray}
On-shell, the last equation (\ref{qj0}) is zero. This essentially
means that $q_\a$ acts as a dimension lowering 
operator such that $\Delta_1=\Delta_2-1/2$ and
$\Delta_0=\Delta_1-1/2$, where $Dj_x=\Delta_xj_x$. Relevant to us is
the case
$\Delta_0=1$ when $j_0$, $j_2$ and $j_1$ are respectively the two
conformally coupled scalars $h_1^{(+)}$ and $h_2^{(-)}$ and the bulk spinor $\Psi$
in (\ref{bulkaction}) Again, ${\cal N}=1$ SUSY requires that one assigns different parities
to the two scalar components of the multiplet. This can be easily
seen if one realizes from (\ref{qqP}) and (\ref{qj1}) that the
operator that lowers the dimension of $j_2$ by unity is proportional
to $\epsilon^{\a\b}q_\a q_\b$ which is odd under parity \cite{Nicolai}. On the spinor, parity
acts as in (\ref{parity1}) where now $\Pi=\Gamma^2$. A mass term on
AdS$_4$ is parity odd.

\end{appendix}


\end{document}